\documentclass[twocolumn,floatfix,prb,aps,showpacs]{revtex4-1}
\usepackage{graphicx,epsf,amsmath,amssymb}
\usepackage{bm,bbm}
\usepackage{multirow}

\usepackage{color}

\newcommand{\BEDT}{$\alpha$-(BEDT-TTF)$_2$I$_3$}

\newcommand{\bone}{\mathbbm{1}}

\newcommand{\sigmab}{\mbox{\boldmath $\sigma $}}

\newcommand{\Rmath}{\mathcal{R}}
\newcommand{\bp}{{\bf p}}

\newcommand{\br}{{\bf r}}
\newcommand{\be}{{\bf e}}

\newcommand{\bA}{{\bf A}}

\newcommand{\beq}{\begin{equation}}
\newcommand{\beqn}{\begin{eqnarray}}
\newcommand{\eeq}{\end{equation}}
\newcommand{\eeqn}{\end{eqnarray}}

\begin{document}

\title{Magneto-optics of quasi-relativistic electrons in graphene with an inplane electric field and in tilted Dirac cones in \BEDT}

\author{Judit S\'ari$^{1,2}$, Mark O. Goerbig$^3$, and Csaba T\H oke$^{2}$}
\affiliation{$^{1}$Institute of Physics, University of P\'ecs, H-7624 P\'ecs, Hungary}
\affiliation{$^{2}$BME-MTA Exotic Quantum Phases ``Lend\"ulet" Research Group,
Budapest University of Technology and Economics,
Institute of Physics, Budafoki \'ut 8, H-1111 Budapest, Hungary}
\affiliation{$^{3}$Laboratoire de Physique des Solides, CNRS UMR 8502, Universit\'e Paris-Sud, F-91405 Orsay Cedex, France}
\date{\today}

\begin{abstract}
Massless Dirac fermions occur as low-energy modes in several quasi-two-dimensional condensed matter systems
such as graphene, the surface of bulk topological insulators, and in layered organic semiconductors.
When the rotational symmetry in such systems is reduced either by an in-plane electric field or an intrinsic
tilt of the Dirac cones, the allowed dipolar optical transitions evolve from a few selected transitions into a wide fan of
interband transitions.
We show that the Lorentz covariance of the low-energy carriers allows for a concise analysis of
the emerging magneto-optical properties.
We predict that infrared absorption spectra yield quantitative information on the tilted Dirac cone structure in
organic compounds such as \BEDT.
\end{abstract}


\maketitle

\section{Introduction}

The low-energy carriers in graphene imitate the relativistic physics of massless Dirac fermions
with the Fermi velocity $v\approx c/300$ replacing the speed of light $c$.\cite{graphenedirac}
Their peculiar Landau level (LL) structure provided the first evidence for the successful fabrication
of monolayer graphene.\cite{monolayer}
Experimental progress in the last decade has allowed, for example,
the direct measurement of the Dirac cones in photoemission experiments,\cite{basovRev}
the observation of LLs in magneto-optical \cite{MO} and scanning-tunneling experiments,\cite{STS}
and the detection of Klein tunneling.\cite{KleinT}
Further predictions follow from the Lorentz covariance of low-energy electrons in graphene,
notably the collapse of Landau levels subjected to in-plane electric and perpendicular magnetic fields.\cite{lukose}
Here, conventional Landau states exist only as long as a Lorentz boost is possible to a reference frame
where the external electric field vanishes;
both the spectrum and the eigenstates follow by Lorentz covariance.
Similar ideas apply to massless Dirac surfaces states of bulk topological insulators.\cite{TI}
On the theory side, exploiting the Lorentz covariance of the massless Dirac equation often yields
very concise derivations (compare, e.g., Refs.~\onlinecite{lukose} and \onlinecite{Peres}),
with the additional advantage of a unified understanding of a class of high-energy and low-energy phanomena.

\begin{figure}
\epsfysize+6.5cm
\epsffile{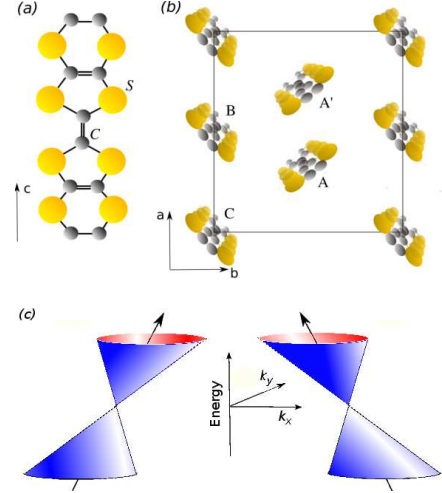}
\caption{(Color online)
Organic compound \BEDT. (a) Molecule (BEDT-TTF) that consists of carbon (C) and sulfur (S) atoms. 
(b) Arrangement of the four molecules (A, A', B, and C) in the unit cell of a single layer ($ab$-plane). 
(c) Sketch of the low-energy band structure in the vicinity of the Fermi level that has the form of two tilted Dirac cones.
We set up the coordinate system so that the cones are tilted in the $k_x$ direction.}
\label{fig01}
\end{figure}

More recently, the quasi-2D organic compound \BEDT\ has been shown to host Dirac cones at low
energies\cite{katayama,kobayashi} under pressure \cite{pressure} or uniaxial strain.\cite{strain}
Other quasi-2D organic materials that may share the same property, e.g.,
$\theta-$(BEDT-TTF)$_2$I$_3$ \cite{theta,theta2} and
$\alpha$-(BEDT-TTF)$_2$NH$_4$Hg(SCN)$_4$,\cite{Choji2011} still need further investigation.
In contrast to graphene, the Dirac cones in these materials are tilted (see Fig.~\ref{fig01}).
We will refer to such compounds as \textit{2D Weyl materials}.
In a magnetic field, this tilt plays the role of an effective in-plane electric field\cite{goerbigEPL}
and can be treated, as we show here, in a covariant manner.
Moreover, the covariance of Dirac electrons in graphene and in \BEDT\ yields another relativistic effect that consists of 
highly unusual magneto-optical transitions. 
Whereas in the frame of reference, where the electric field (or the tilt in 2D Weyl materials) vanishes, one obtains the usual 
dipolar transitions between the LLs $\pm n$ and $\pm(n\pm 1)$,\cite{selection} we predict that
a Lorentz boost back to the laboratory frame gives rise to an amazing multiplication of these optical lines into a fan of experimentally 
observable interband transitions.
Finally, we show that magneto-optical measurements may yield information on the tilt of the Dirac cones
in 2D Weyl materials.

This article is structured as follows.
In Sec.~\ref{models} we review the low-energy band structure of the materials we study
together with the relevant single-particle (Landau) states.
In Sec.~\ref{intera} we study the matrix elements of the interaction of the electrons with electromagnetic radiation.
The optical absorption per layer is analyzed in Sec.~\ref{abso}, where we also discuss how information can be gathered
on the tilted Dirac cone structure of 2D Weyl materials from magneto-optics.
In Sec.~\ref{conclu} we conclude and discuss the experimental connections.
Details of the calculation are delegated to the Appendices.

\section{Materials and models}
\label{models}

Let us first discuss graphene electrons in a perpendicular magnetic field $\mathbf B=B_\perp\mathbf e_z$
and in-plane electric field $\mathbf E=-E_\parallel\mathbf e_y$.
The low-energy carriers are described by the Hamiltonian\cite{graphenedirac}
\begin{equation}
\label{grhami}
\hat{H}_\text{G}= v(\bp + e\bA)\cdot\sigmab - eE_\parallel y\bone,
\end{equation}
where $v$ is the Fermi velocity of graphene, the Pauli matrices are denoted by $\sigma_i$ with $i\in\{x, y, z\}$
and $\sigmab = (\sigma_x, \sigma_y)$.
Hereafter, we use the Landau gauge $\bA=-yB_\perp \be_x$ and $e>0$.
With the choice of $x^\mu =(vt,x,y)$ in the laboratory frame $\Rmath$ one may introduce the 2D massless Dirac equation
\begin{equation}
\label{lorentz}
i\hbar v \gamma_\mu \left(\partial^\mu- \frac{i e}{\hbar}A^\mu\right) \Psi=0,
\end{equation}
where the representation $\gamma_0=\sigma_z$, $\gamma_1=-\sigma_z\sigma_x$, $\gamma_2=-\sigma_z\sigma_y$ has been used,
and $A^\mu=(E_\parallel y/v,-yB_\perp,0)$.
The electric field can be eliminated by a Lorentz boost in the $x$-direction from the laboratory frame $\Rmath$ to
$(vt',x',y')=\Lambda(vt,x,y)=(\gamma (v_Dt + \beta x),\gamma (v_D t + x),y)$ 
to another reference frame $\Rmath'$, as long as the drift velocity $v_D=E_\parallel/B_\perp$ is smaller
than the Fermi velocity, $v_D<v$. The Lorentz boost is thus parametrized by the 
rapidity $\tanh\theta\equiv\beta=v_D/v$ and the Lorentz factor $\gamma = 1/\sqrt{1-\beta^2}$.
The Hamiltonian in $\Rmath'$ then reads 
\begin{eqnarray}
\label{eq:ham}
\hat{H}_\text{G}'&=&v(\bp'+e\bA')\cdot\sigmab,
\end{eqnarray}
and  covariance requires the two-spinor $\Psi$ to transform as 
$$\Psi'(vt',x',y')=S(\Lambda)\Psi(vt,x,y),$$
with $S(\Lambda)= \exp(\frac{\theta}{2}\sigma_x)$.

Notice that the simple form of $\hat{H}_\text{G}'$ allows for a straightforward diagonalization in $\Rmath'$ by standard techniques.\cite{goerbigRev,lukose,Peres}
The sought eigenstates of $\hat{H}_\text{G}$ are then obtained by a transformation back to $\Rmath$,
\begin{multline}\label{eq:WFn}
\Psi_{\lambda,n;k}(x,y) =\frac{1}{\sqrt{2}}\left[
\left(\begin{array}{c} \sinh\frac{\theta}{2} \\ -\cosh\frac{\theta}{2} \end{array}\right)\phi_{\lambda,n;k}(x,y)\right.\\
\left.+ \lambda \left(\begin{array}{c} \cosh\frac{\theta}{2} \\ -\sinh\frac{\theta}{2} \end{array}\right)\phi_{\lambda,n-1;k}(x,y)
\right]
\end{multline}
for $n\geq 1$ and
\beq\label{eq:WF0}
\Psi_{n=0;k}(x,y)=\left(\begin{array}{c} \sinh\frac{\theta}{2} \\ -\cosh\frac{\theta}{2} \end{array}\right)\phi_{n=0;k}(x,y)
\eeq
for $n=0$, where $\lambda=\pm$ denotes LLs of positive and negative energy, respectively, and $k$ is the wavevector in the $x$-direction related
to the conserved momentum component $p_x=\hbar k$.
Here we have used the functions
\begin{gather}
\label{eq:wave}
\phi_{\lambda,n;k}(x,y)=\frac{e^{ikx}}{\sqrt{2\pi}}\frac{1}{\sqrt{\sqrt{\pi\gamma}\ell 2^n n!}}
H_n\left(\eta_{\lambda,n}\right)e^{-\eta_{\lambda,n}^2/2},\\
\eta_{\lambda,n}=\frac{1}{\sqrt{\gamma}\ell}\left(y - k\ell^2\right) - \lambda\beta\sqrt{2n},
\end{gather}
where $\ell=\sqrt{\hbar/eB}$ is the magnetic length, $H_n$ is a Hermite polynomial,
and $\delta$-normalization has been used for the wave functions.
The covariant character is apparent here in the argument $\eta_{\lambda,n}$ of the wavefunctions,
where one notices the contraction factor $1/\sqrt{\gamma}$.
The second term reflects the admixing of the time-like component, $\lambda \sqrt{2n}$, proportional to the LLs,
\beq\label{eq:grLL}
\epsilon^\text{G}_{\lambda,n;k}=\frac{\lambda \hbar v}{\gamma^{3/2} \ell}\sqrt{2n} -\hbar v_D k,
\eeq
which exhibit a collapse in the $v_D\to v$ ($\gamma\to\infty$) limit.\cite{lukose,Peres}
The last term in Eq.~(\ref{eq:grLL}) stands for the dependence of the single-particle energies
on the location of the guiding center if an in-plane electric field is present.
This term is also present in the case of nonrelativistic electrons, but the LL spacing of the latter is insensitive to the inplane
electric field, in contrast to the decreased LL spacing of relativistic electrons apparent in the factor $1/\gamma^{3/2}$ in Eq. (\ref{eq:grLL}).

The low-energy carriers in 2D Weyl materials are described by the minimal Weyl
Hamiltonian,\cite{kobayashi,Morinari,Goerbig}
which allows for different velocities $v_x$, $v_y$ in two directions, and a shift of the centers of the
equienergy ellipses in a generic tilt direction.
The anisotropy $v_x\neq v_y$ can be eliminated by rescaling $y'=(v_x/v_y)y$,
and rotating the direction of the tilt into the $x$-direction.
The details of this procedure are presented in Appendix \ref{derivation}.
We obtain
\beq\label{eq:HamOrg}
\hat{H}_\text{W}=v_x(\bp+e\bA)\cdot\sigmab + v_0(p_x + eA_x)\bone.
\eeq
Here $v_0$ is a velocity parameter that is related to the band structure, c.f.\ Eq.~(\ref{vnull}).

In the Landau gauge the noncovariant term $v_0p_x\bone$ in Eq.~(\ref{eq:HamOrg})
is harmless as $p_x$ only extracts a quantum number.
The Hamiltonian (\ref{eq:HamOrg}) can be rewritten in 
the form $\hat{H}_\text{W}=\hat{H}_\text{W}^\text{cov.} + v_0p_x\bone$,
where the last term is diagonal
both with respect to the sublattice $\sigmab$ and in the wave vector $k$.
The energy levels can therefore be written as 
$$
\epsilon^\text{W}_{\lambda,n;k} = \epsilon^\text{cov.}_{\lambda,n;k} + \hbar v_0 k,
$$
where $\epsilon^\text{cov.}_{\lambda,n;k}$ are the solutions of the Hamiltonian's
covariant part $\hat{H}_\text{W}^\text{cov.}$, which has the same
form as that in Eq.~(\ref{grhami}).
In the Landau gauge, the term $ev_0A_x\bone=-ev_0B_{\perp}y\bone$ can be viewed as
generated by an effective pseudo-electric field $E^{\rm eff}_\parallel=B_\perp v_0$,
and $v_0$ assumes the role of the drift velocity.
The wave functions are obtained by methods that are completely analogous to the graphene case,
using $\tanh\theta\equiv\beta=v_0/v$ in the Lorentz boost.
Then the LL spectrum becomes\cite{Goerbig,Morinari}
\beq\label{eq:weylLL}
\epsilon^\text{W}_{\lambda,n;k}=\frac{\lambda \hbar v}{\gamma^{3/2} \ell}\sqrt{2n}.
\eeq
where an average velocity $v=\sqrt{v_xv_y}$ has been introduced.
Comparing with Eq.~(\ref{eq:grLL}), the guiding-center-dependent term is absent, because there is
no inplane electric field.

Both in graphene and in 2D Weyl materials, Dirac fermions occur in two cones related by time-reversal symmetry.
Whereas electrons in both valleys are subjected to the same external electric field (drift velocity),
the tilt in Weyl materials is opposite in the two valleys and so is the implied effective electric field.

\section{Interaction with electro-magnetic radiation}
\label{intera}

Let the 2D sample be located in the $z=0$ plane.
We consider the interaction with a classical radiation field
\begin{equation}
\label{Ateljes}
\mathbf A^\text{rad}(\mathbf r,t)=\mathrm{Re}\left[\frac{E_0}{\omega}
\left(\alpha\mathbf{\hat e}_x+\tau\mathbf{\hat e}_y\right)e^{i \omega t}\right],
\end{equation}
where $|\alpha|^2+|\tau|^2=1$ and $\omega$ is fixed;
$\bA^\text{rad}(\br,t)$ is added to the vector potential in Eqs.~(\ref{grhami}) and (\ref{eq:HamOrg}).
Notice that the different single-particle Hamiltonians for graphene and 2D Weyl materials
yield slightly different perturbation operators
\begin{equation}
\label{pertHG}
^\text{G}\delta \hat H=
e v \frac{E_0}{\omega} \left(\alpha \sigma_x + \tau \sigma_y\right)
\end{equation}
and
\begin{equation}
\label{pertHW}
^\text{W}\delta \hat H=
e v_x  \frac{E_0}{\omega}  \left(\tilde{\alpha} \sigma_x +  \tilde{\tau} \sigma_y + \frac{v_0}{v_x} \tilde{\alpha} \bone\right),
\end{equation}
respectively.
The parameters $\alpha$ and $\tau$ determine the light polarization, e.g.,
$\alpha=1/\sqrt{2}$ and $\tau=\mp i/\sqrt{2}$ for circular polarization $\circlearrowright/\circlearrowleft$.
In the case of 2D Weyl materials, $\alpha$ and $\tau$ are also affected by the rescaling of the coordinates
(c.f.\ Appendix \ref{derivation}); the use of tildes for $\tilde\alpha$ and  $\tilde\tau$ refer to this change.

\begin{figure}[htbp]
\begin{center}
\includegraphics[width=0.46\columnwidth,keepaspectratio]{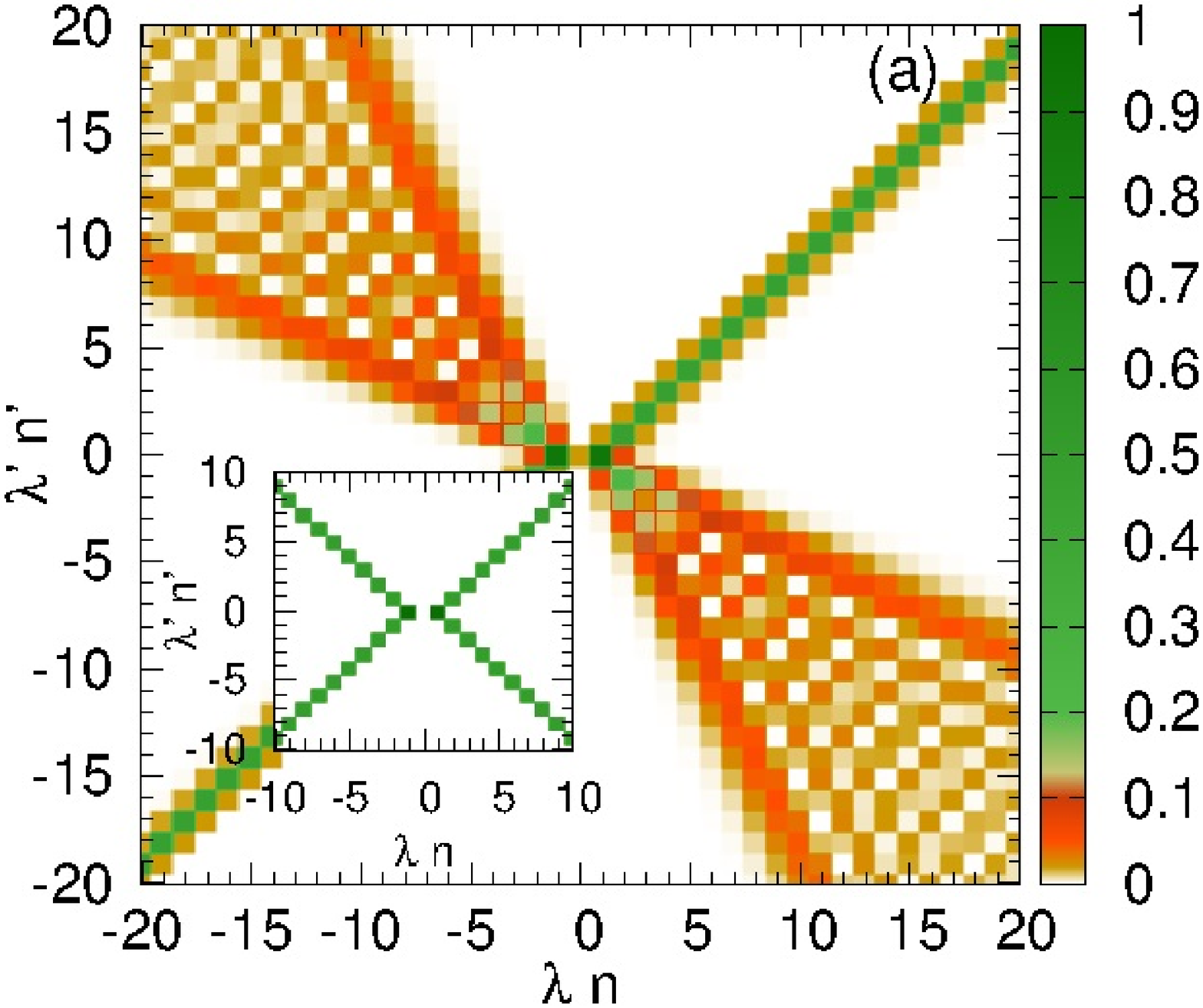}
\includegraphics[width=0.47\columnwidth,keepaspectratio]{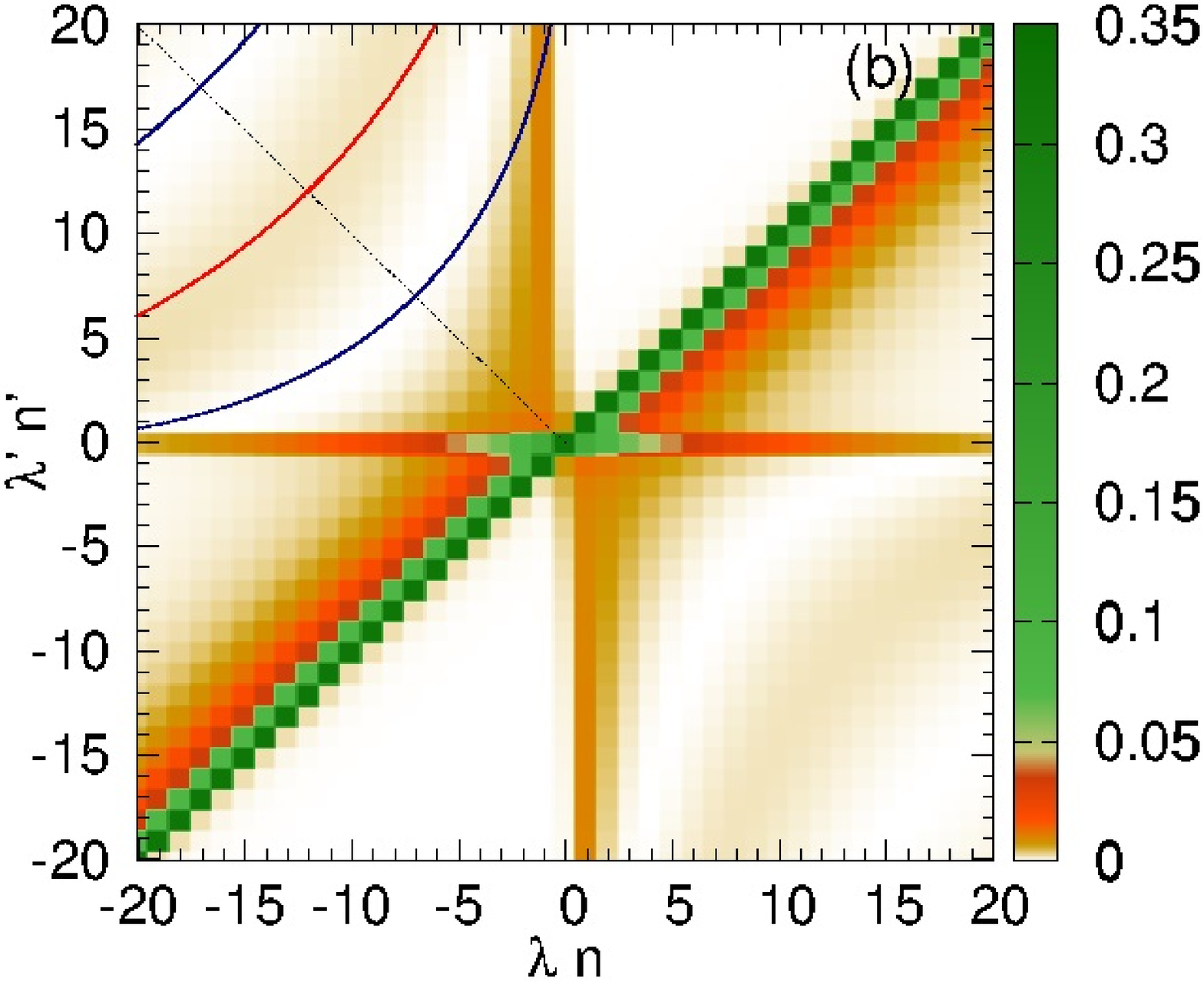}

\includegraphics[width=0.46\columnwidth,keepaspectratio]{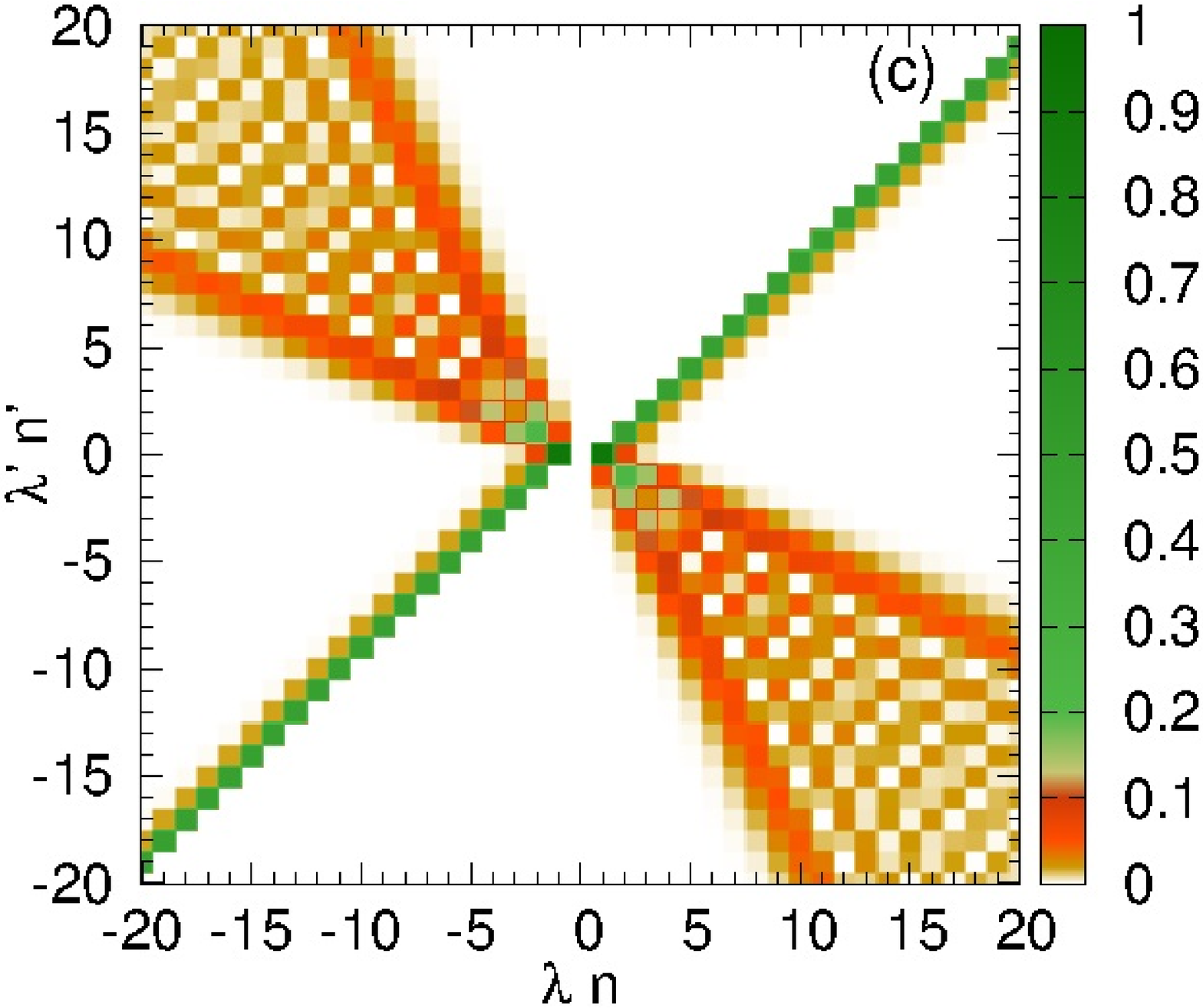}
\includegraphics[width=0.47\columnwidth,keepaspectratio]{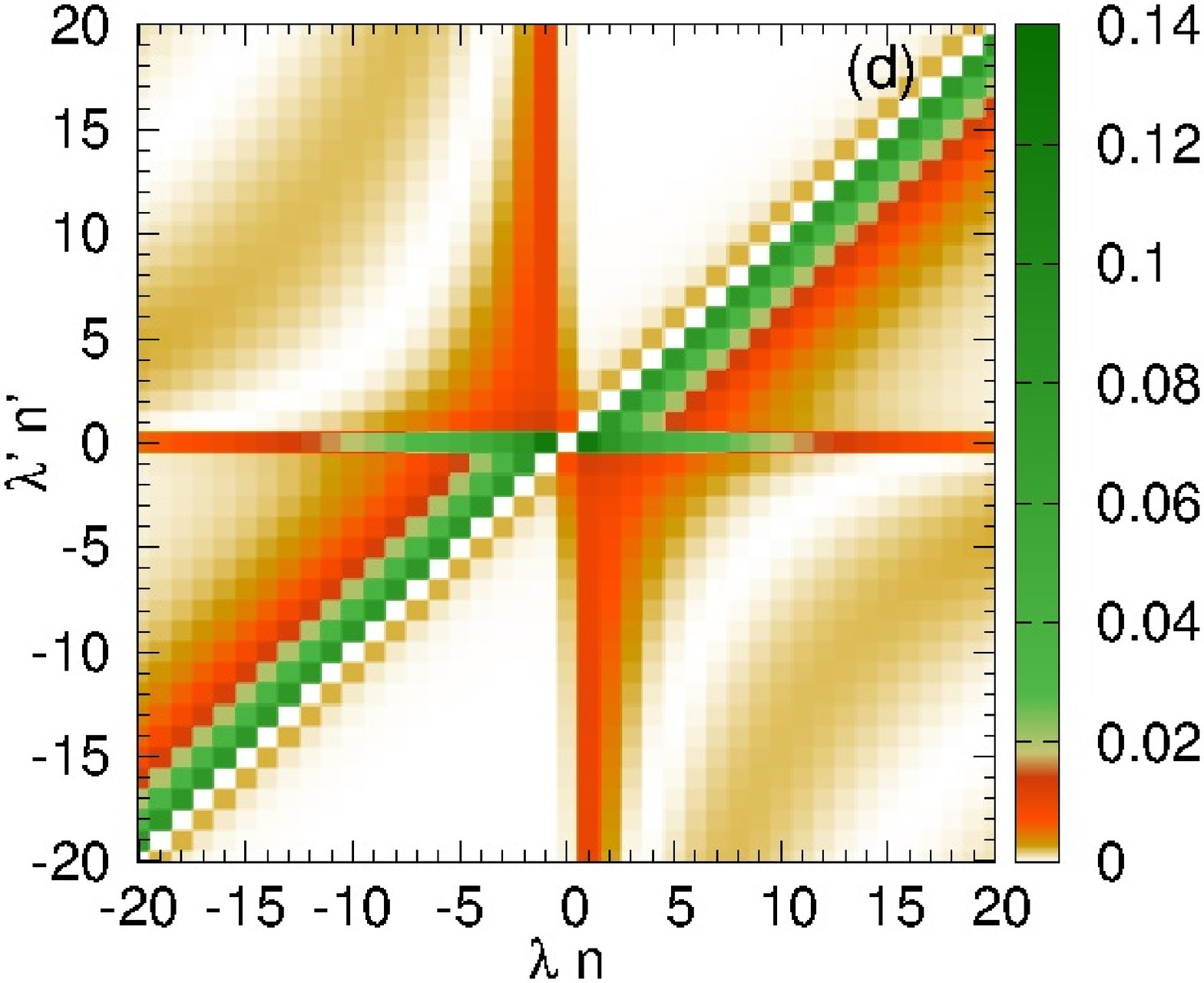}
\end{center}
\caption{\label{nmtable}
(Color online)
(a,b) The matrix elements $|^\text{G}Q^{\lambda',n'}_{\lambda,n}|^2$
for graphene in crossed $B_\perp$ and $E_\parallel$ fields, for right circularly polarized light.
(c,d) $|^\text{W}Q^{\lambda',n'}_{\lambda,n}|^2$
for a 2D Weyl material, for elliptical polarization that appears as right circularly polarized in the rotated and rescaled system.
(a,c) $\beta=0.2$ and (b,d) $\beta = 0.8$.
Inset of panel (a): the same as in (a) for $E_\parallel = 0$ limit.
In panel (b) the dotted line indicates the matrix elements corresponding to $n'=n\ge1,\lambda'=-\lambda$ transitions, see Eq.~(\ref{diagonal}); blue (red) curves indicate equi-energy lines with minimal (maximal) value of the matrix elements.
}
\end{figure}

For the electron-light interaction, we will need the matrix elements of $\delta\hat H^\text{G/W}$:
\begin{equation}
\label{matelem}
^\text{G/W}\delta H^{\lambda',n',k'}_{\lambda,n,k}=
e v \frac{E_0}{ \omega } \delta (k - k')\left( {^\text{G/W}Q}^{\lambda',n'}_{\lambda,n}\right),
\end{equation}
where the dimensionless factors ${^\text{G}Q}^{\lambda',n'}_{\lambda,n}$ and ${^\text{W}Q}^{\lambda',n'}_{\lambda,n}$
of these matrix elements are
\begin{multline}
{^\text{G}Q}^{\lambda',n'}_{\lambda,n}=
-\frac{1}{2}\left[ F_{n, n'}^{\lambda, n,\lambda', n'} \alpha \beta +
\lambda'\lambda F_{n-1, n'-1}^{\lambda, n,\lambda', n'} \alpha\beta\right.\\ 
\left.+ \lambda' F_{n,n'-1}^{\lambda, n,\lambda', n'} (\alpha + i \tau \gamma^{-1} ) +
\lambda F_{n-1,n'}^{\lambda, n,\lambda', n'} (\alpha - i \tau \gamma^{-1} ) \right],
\end{multline}
and
\begin{multline}
{^\text{W}Q}^{\lambda',n'}_{\lambda,n}=
\frac{1}{2\gamma}\left[  \lambda' F_{n,n'-1}^{\lambda, n,\lambda', n'} (\gamma^{-1}  \tilde \alpha - i \tilde \tau  )\right.\\
\left.+\lambda F_{n-1,n'}^{\lambda, n,\lambda', n'}  (\gamma^{-1} \tilde \alpha + i \tilde \tau ) \right].
\end{multline}
The functions $F^{\lambda, s, \lambda' , p}_{q,r}$ are defined in terms of associated Laguerre polynomials $L_n^m$, as
\begin{multline}
\label{auxfunction}
F^{ \lambda, s, \lambda', p}_{q, r}=\sqrt{\frac{r!}{q!} } e^{-\left(z_{\lambda, s}^{\lambda', p}\right)^2 / 2}
\left(z_{\lambda, s}^{\lambda', p}\right)^{q -r} \ L_{r}^{q -r}\left[\left(z_{\lambda, s}^{\lambda', p}\right)^2 \right],\\
z_{\lambda, s}^{\lambda', p} = \beta (\lambda' \sqrt{p} -  \lambda \sqrt{s}),
\end{multline}
for $q \geq r$, otherwise $F^{ \lambda, s,\lambda', p}_{q,r}=F^{\lambda', p, \lambda, s}_{r,q}$.

We show $|^\text{G/W}Q^{\lambda',n'}_{\lambda,n}|^2$ in Fig.~\ref{nmtable} for graphene and 2D Weyl materials
for $\beta=0.2$ and $0.8$, assuming right circular polarization and a special choice of elliptical polarization,
respectively.
The latter choice is determined by the asymmetric velocities and corresponds to a circular polarization after rescaling.
The opposite polarization is obtained via the relation
\begin{equation}
\label{symmq}
|(Q^\circlearrowleft)^{\lambda',n'}_{\lambda,n}|^2=|(Q^\circlearrowright)^{\lambda',n}_{\lambda,n'}|^2,
\end{equation}
and both $|(Q^{\circlearrowleft/})^{\lambda',n'}_{\lambda,n}|^2$ and
$|(Q^{\circlearrowright})^{\lambda',n'}_{\lambda,n}|^2$ are independent of the sign of $\beta$.
The results for both systems are almost identical, apart from small differences due to the slightly different
perturbation operators in Eq.~(\ref{pertHG}) and (\ref{pertHW}).
Most saliently, we obtain novel transitions beyond the usual dipolar ones, $\propto\delta_{n',n\pm1}$
that arise for vanishing $E_\parallel$ or no tilt, respectively, and
which apply therefore only in the reference frame $\Rmath'$, where rotation symmetry is restored. Alternatively,
one may thus view the occurrence of additional transitions as due to the broken rotation symmetry caused by the 
Lorentz boost back to the laboratory frame $\Rmath$.
The weight of \textit{intraband} transitions that change 
the LL index $n$ by more than one unit grows monotonically with $\beta$ whereas
the \textit{interband} transitions show a complex fan structure with an arc pattern. 
The fan, whose opening angle increases with $\beta$,
is bounded by large values of the perturbation matrix element that are not affected by the arc pattern
[c.f.\ also the inset of Fig.~\ref{absorption}(b)].

\begin{figure}[htbp]
\begin{center}
\includegraphics[height=\columnwidth,angle=270,keepaspectratio]{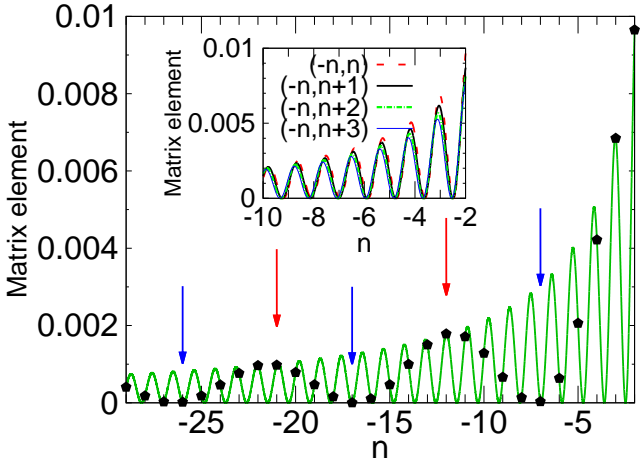}
\end{center}
\caption{\label{wavy}
(Color online)
The matrix element $|^\text{G}Q^{-\lambda,n}_{\lambda,n}|^2$ as a continuous function of $n$ (green) for $\beta=0.8$.
The black dots indicate integer values of $n$.
Inset: the same along the lines $\lambda'=-\lambda$ and $n'=n+c$, where $c=1,2,3$.
}
\end{figure}

\begin{figure}[htbp]
\begin{center}
\includegraphics[height=\columnwidth,angle=270,keepaspectratio]{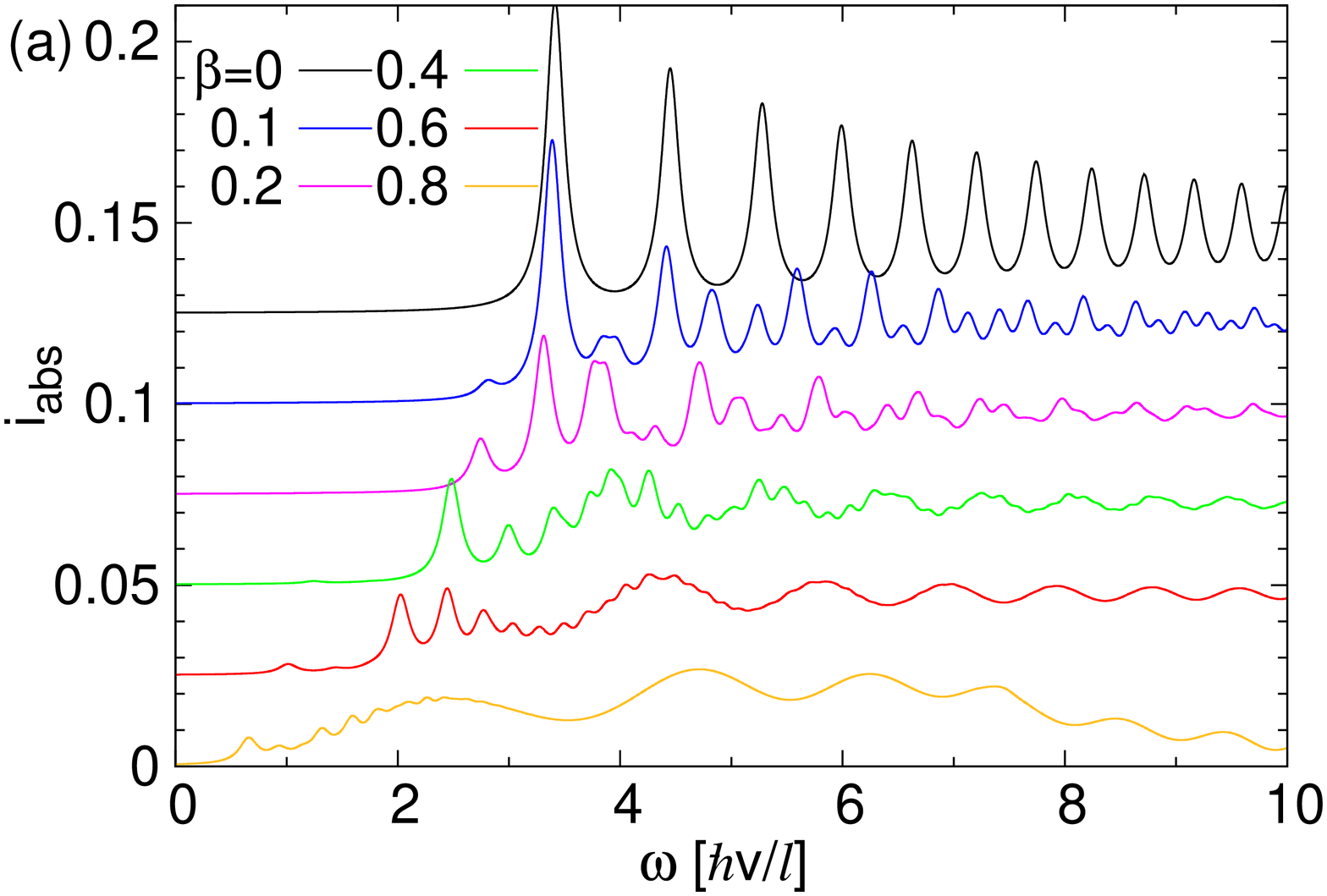}
\includegraphics[height=\columnwidth,angle=270,keepaspectratio]{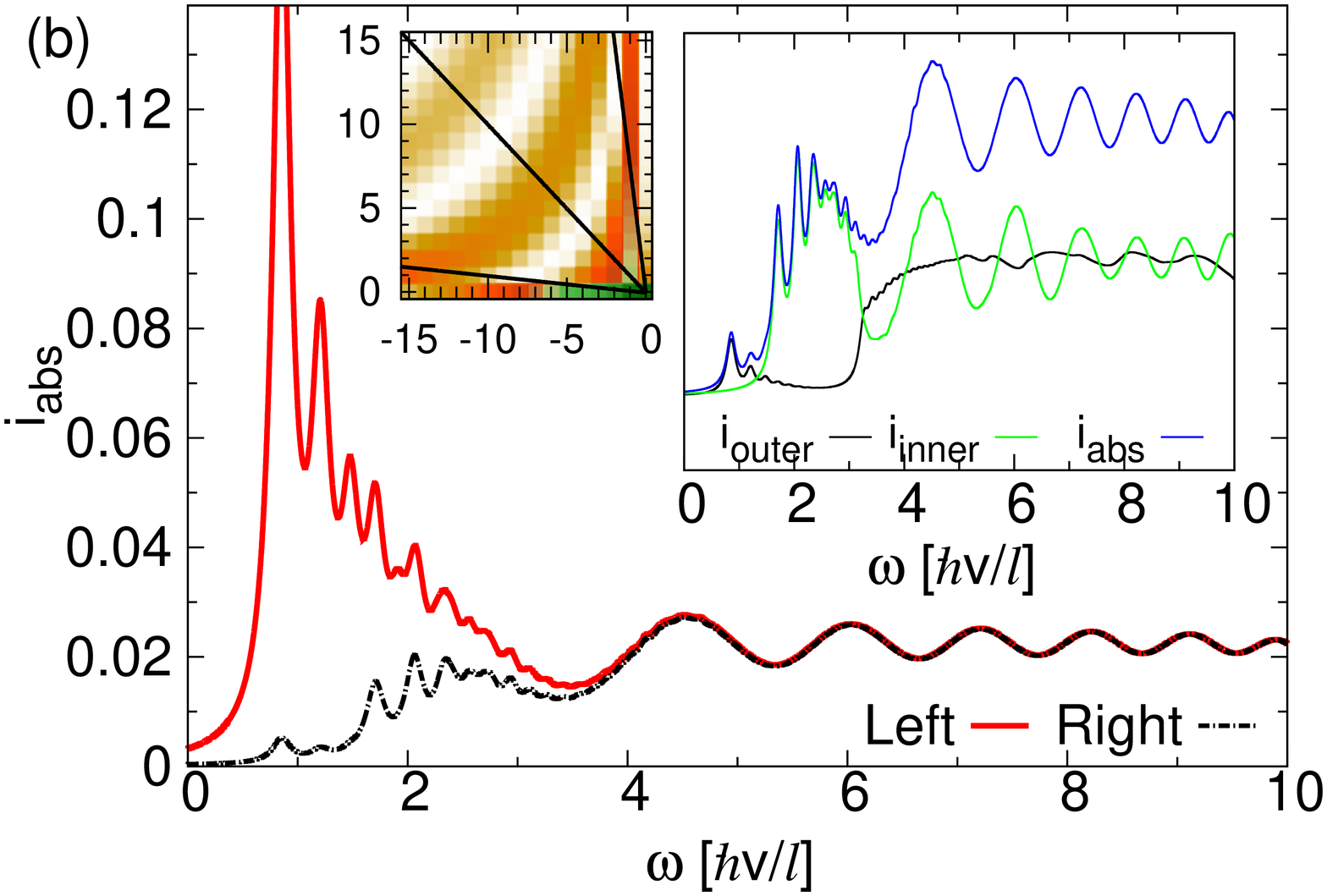}
\end{center}
\caption{\label{absorption}
(Color online)
(a) Absorption coefficient of graphene in perpendicular magnetic and in-plane electric fields
for right circularly polarized light.
Curves are shifted for clarity.
(b) Comparison of $i_\text{abs}$ for right and left circularly polarized light for $\beta = 0.7$.
Inset: The total $i_\text{abs}$, $i_\text{inner}$ due to the inner part of the interband fan, and $i_\text{outer}$ due to the outer part,  for graphene $\beta = 0.7$.
}
\end{figure}

For further insight into the origin of the observed arc pattern,
consider the case $n'=n\ge1,\lambda'=-\lambda$ for large $\beta$.
One finds that
\begin{equation}
^\text{G/W}Q^{-\lambda,n}_{\lambda,n}
\propto e^{-2\beta^2n}L^1_{n-1}(4\beta^2n),
\label{diagonal}
\end{equation}
where the associated Laguerre polynomial $L_n^m$
can be generalized to continuous $n$ in terms of the confluent hypergeometric function.
At fixed $\beta$, this function has decaying oscillations in $n$.
For $\beta= 0.8$, this function is plotted in Fig.~\ref{wavy}(a), where we have used $n$ as a continuous variable.
The apparent period $\delta n$ is more than, but close to, unity, and this incongruity
with integer values yields a beating with a distance $\Delta n\approx10$ between consecutive zeros that gives rise to the observed arc pattern.
By Nyquist's sampling theorem, this effect survives as long as $\delta n<2$ (which corresponds to $\beta\gtrsim0.4$).
Nearby transitions such as $(\lambda',n')=(-\lambda,n+1)$, $(\lambda',n')=(-\lambda,n+2)$, etc.,
can be analyzed similarly.
Remarkably, their beatings are of almost identical period and are in phase (Fig.~\ref{wavy} inset).
While the latter phase condition does not hold for small $\beta$ in general, 
we will see below that two characteristic periods can be extracted, one of which is continuously
connected to the period at large $\beta$.

\section{Magneto-optical absorption}
\label{abso}

By Fermi's Golden Rule,\cite{Zettili} the transition rates per unit area $A$ are
\begin{multline}
R_{\lambda, n; k \to \lambda' , n';k'}^\text{emi/abs} =
\frac{1}{A} \frac{2 \pi}{\hbar}
\left| ^\text{G/W}\delta H^{\lambda',n',k'}_{\lambda,n,k}\right|^2\\
\times\delta(\epsilon_{\lambda', n'} - \epsilon_{\lambda , n} \pm \hbar \omega)
n_F(\epsilon_{\lambda, n} ) [1- n_F(\epsilon_{\lambda' , n'})],
\end{multline}
where $n_F(\epsilon)$ is the Fermi function.
Integrating over all states, the total transition rate per unit area is
\begin{equation}
R = \sum_{\lambda, n, \lambda', n' } \int \mathrm{d} k\, \int \mathrm{d} k'
\left[R_{\lambda, n; k \to \lambda' , n';k'}^\text{abs} - R_{\lambda', n'; k' \to \lambda , n; k}^\text{emi}\right].
\end{equation}
This can be normalized using the absorbed/emitted energies and the magnitude of the Poynting vector $S = E_0^2 / 2\mu_0 c$,
\begin{equation}
i_\text{abs}^\text{G/W}(\omega) = \frac{R \hbar \omega}{S},
\end{equation}
which is dimensionless.
(The celebrated 2.3\% per sheet absorbtion of pristine graphene\cite{sheet} employs this definition.)
We introduce a Lorentzian broadening $\Gamma$ for neglected effects (temperature, impurities, phonons, etc).
Taking spin and valley into account, one obtains
\begin{multline}
i_\text{abs}^\text{G/W}(\omega) = \frac{2 v^2\hbar\mu_0ce^2}{\ell^2}\sum_{\lambda,n, \lambda',n'}
\frac{n_\text{F}(\epsilon_{\lambda,n})-n_\text{F}(\epsilon_{\lambda',n'})}
{\epsilon_{\lambda',n'} - \epsilon_{\lambda,n}}\\
\times\left|^\text{G/W}Q^{\lambda',n'}_{\lambda,n}\right|^2
\frac{1}{\pi}\frac{\Gamma}{\Gamma^2 + (\epsilon_{\lambda',n'} - \epsilon_{\lambda,n} - \hbar\omega)^2}.
\end{multline}
Inspecting the graphene data by Sadowski \textit{et al.}\cite{MO} we will use
$\Gamma=0.1\hbar v/\ell$ as a rough estimation.
Eq.~(\ref{symmq}) implies that the absorption for opposite polarizations are connected
\begin{equation}
i_\text{abs}^{\text{G/W},\circlearrowleft}(\omega, \mu) =
i_\text{abs}^{\text{G/W},\circlearrowright}(\omega, -\mu),
\end{equation}
where $\mu$ is the chemical potential.

The absorption of graphene in perpendicular megnetic and in-plane electric field
is plotted as a function of frequency in Fig.~\ref{absorption}(a) for several values of $\beta$.
Whereas for low frequencies one finds well-pronounced lines corresponding to individual transitions,
in the large frequency domain the lines of many allowed transitions coalesce.
The resulting supermodulation stems from the arc pattern in Fig.~\ref{nmtable},
where the arcs connect different transitions of nearby energies, see the lines
connecting minima/maxima in Fig.~\ref{nmtable}(b) for illustration.
In Fig.~\ref{absorption}(b) we compare the absorption for the two circular light polarizations.
The low-frequency absorption, where only few LL transitions contribute,  depends strongly on polarization.
In contrast,
the absorption becomes polarization-independent at high frequencies because of the coalescence of
a large number of transitions.
However, the effect of the above-mentioned supermodulation is retained.
Finally, in the inset of Fig.~\ref{absorption}(b) we separate the contributions of the patterned inner
wedge of the fan of interband transitions and of the bounding lines to $i_\text{abs}(\omega)$;
we find that the latter merely yields a constant background.

\begin{figure}[htbp]
\begin{center}
\includegraphics[height=\columnwidth,angle =  270, keepaspectratio]{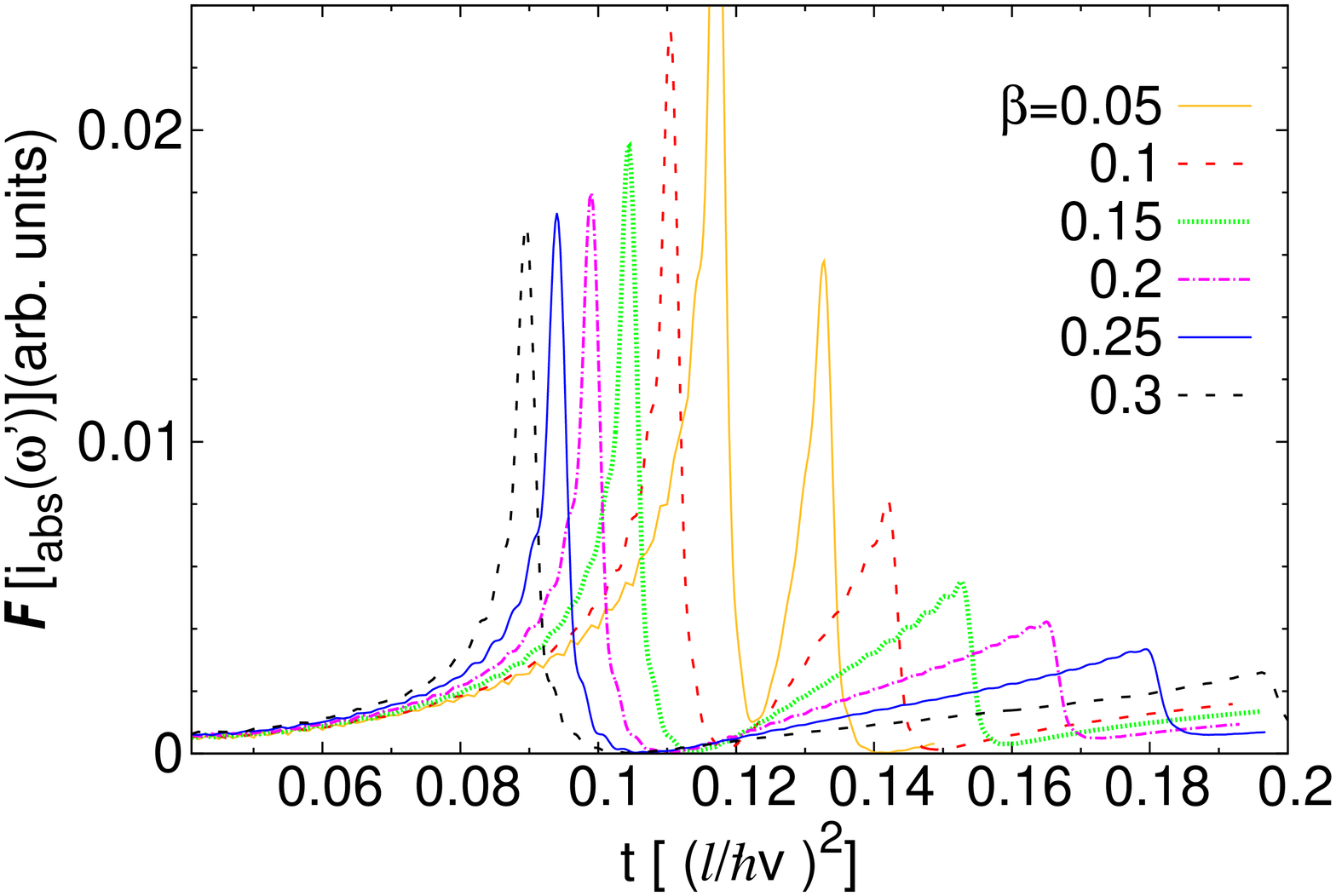}
\includegraphics[height=\columnwidth,angle =  270, keepaspectratio]{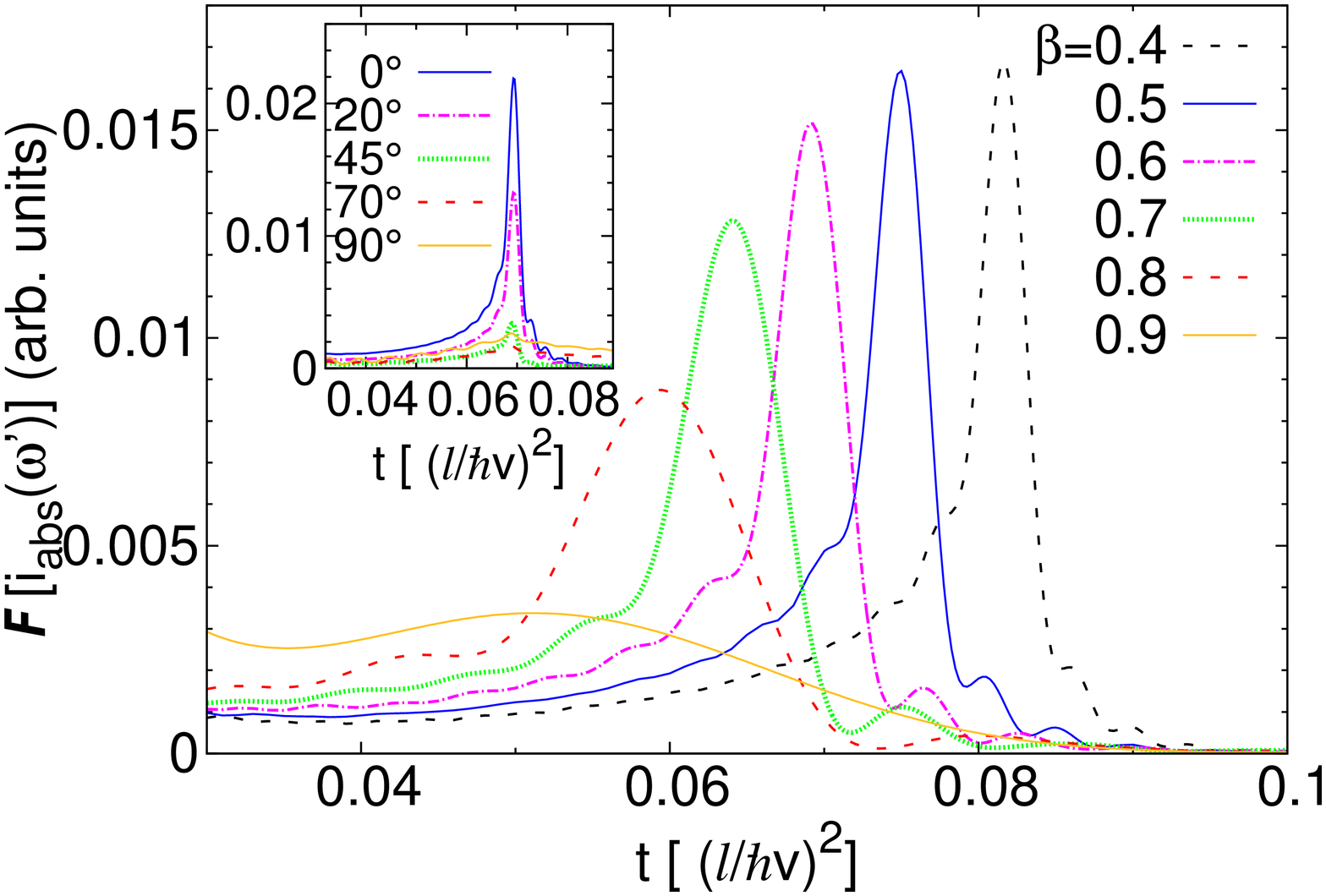}
\end{center}
\caption{\label{cpeak}
(Color online)
(a) The one-sided Fourier-transformed rescaled absorption spectrum $\mathcal{F}[i_\text{abs} (\omega')](t)$
for linearly polarized light exhibits two characteristic peaks in the range $\beta\le0.3$.
We show the case of 2D Weyl materials; the graphene case is similar.
(b) For, $\beta\ge0.4$, only a single characteristic peak is discernible.
Inset: the dependence of the characteristic peak on the direction of linear polarization at $\beta=0.6$,
relative to the tilt direction.
}
\end{figure}

All features in the optical spectra follow the $\epsilon^\text{G/W}_{\lambda,n;k}\propto\sqrt n$ scaling of
massless 2D Dirac fermions.
Therefore, we consider the spectra as a function of $\omega'=\omega^2$, and seek the effect of the
high-frequency regular pattern in their one-sided Fourier transform (or Laplace transform with an imaginary
argument),
\begin{equation}
\label{Ft}
\mathcal{F}[i_\text{abs} (\omega')](t) =
\int_{0}^{\infty} \mathrm{d} \omega' e^{-i 2 \pi t \omega'} i_\text{abs} (\omega=\sqrt{\omega'}).
\end{equation}
A characteristic peak emerges, as shown in the inset of Fig.~\ref{cpeak} for 2D Weyl materials,
even for linearly polarized light.
A particular polarization direction exists for which the visibility is optimal.
For graphene, this is the direction of the electric field;
for 2D Weyl materials, it is the tilt direction.
For small $\beta \lesssim0.35$, a satellite peak appears [c.f.\ Fig.~\ref{cpeak}(a) for the case of
2D Weyl materials], but the main peak remains clearly visible.

\begin{figure}[htbp]
\begin{center}
\includegraphics[height=\columnwidth,angle=270,keepaspectratio]{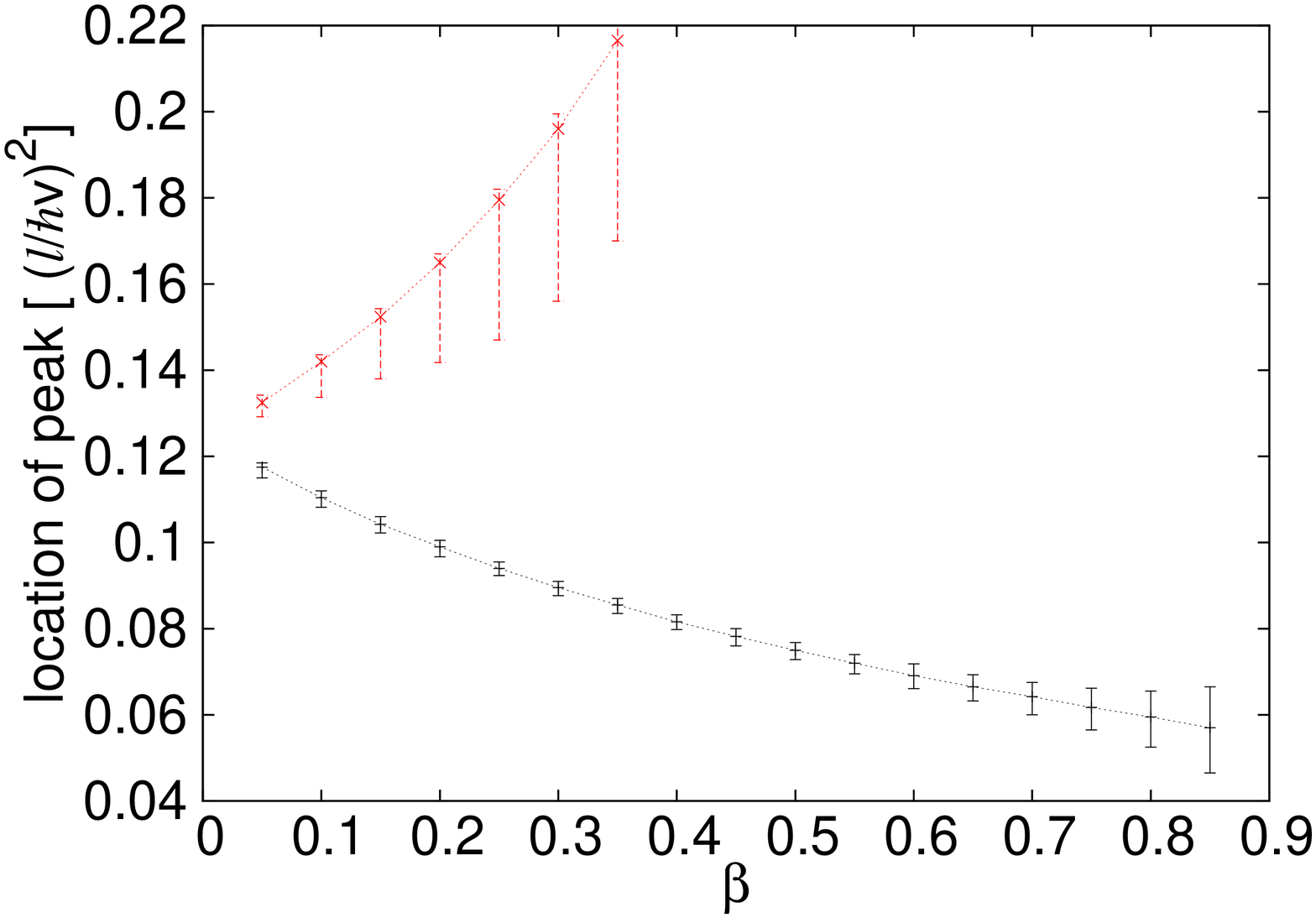}
\includegraphics[height=\columnwidth,angle=270,keepaspectratio]{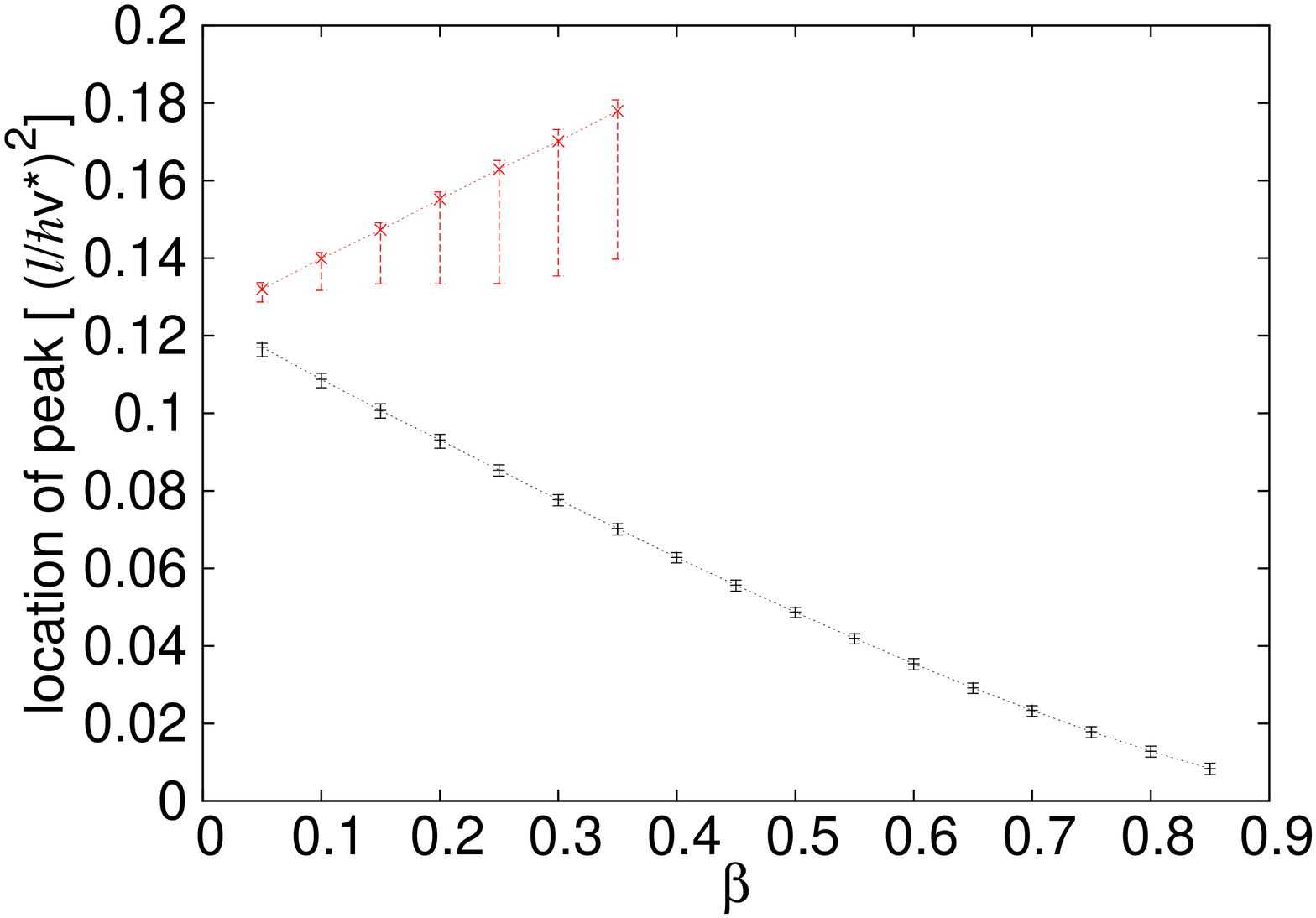}
\end{center}
\caption{\label{graphenelinabra}
(Color online)
(a) The location of the main peak(s) in the one-sided Fourier-transform of the rescaled absorption spectrum
$\mathcal{F}[i_\text{abs} (\omega')](t)$ for linearly polarized light as a function of the tilt
parameter $\beta$ for a 2D Weyl material.
Error bars indicate the FWHM assuming a Lorentzian broadening.
(b) The same data plotted in units of $(\ell/\hbar v^\ast)^2$, where the angle-averaged Fermi velocity
$v^*=\sqrt{v_xv_y}(1-\beta^2)^{3/4}$ is directly related to the density of states that appears in several experiments.
}
\end{figure}

Figure \ref{graphenelinabra} shows the relation between the tilt parameter $\beta$
and the location of the two peaks in the one-sided Fourier-transform of the rescaled spectrum in Eq.~(\ref{Ft}).
Here, we restrict our attention to 2D Weyl materials.
Extracting $\beta$ is straightforward in the moderate-tilt limit, $\beta\lesssim 0.35$,
via the splitting of the two peaks in $\mathcal{F}[i_\text{abs} (\omega')]$.
In the large-tilt region $\beta\gtrsim 0.35$, however, there is only one peak.
Its location is informative, but for quantitative analysis we also need to know the Fermi velocity.
Then it is more useful to plot the location of the peak in units of $(\ell/\hbar v^*)^2$,
where the effective Fermi velocity $v^*=\sqrt{v_xv_y}/\gamma^{3/2}=\sqrt{v_xv_y}(1-\beta^2)^{3/4}$ takes
into account both the anisotropy and the relativistic reduction of the energy scale.
This velocity $v^*$ occurs directly in the density of states,\cite{Morinari,Goerbig}
or in the LL spacing in Eq.~(\ref{eq:weylLL}).
Thus we can obtain the tilt parameter only if we utilize independent information on $v^*$ from thermodynamics
or low-frequency magneto-optics.

\section{Conclusion}
\label{conclu}

In conclusion, the pseudo-relativistic nature of electrons in graphene in crossed electric and magnetic fields, and
in 2D Weyl materials such as \BEDT\ manifests itself in unique magneto-optical properties.
A large number of transitions beyond the usual dipolar ones become possible.
Their coalescence at high frequencies is the fingerprint of the broken rotation invariance in both cases
either due to the particular tilt direction or that of
the electric field.

In graphene, this effect might be observed easier than the earlier predicted Landau level collapse,\cite{lukose}
because the magneto-optical effect is present in the low-$\beta$ range where the change of the Landau level
energies is still small.
A side-gated geometry must be necessary.
Here, however, the screening of the external electric field
by the edge states \cite{edge,edge2} might be a complication.\cite{foot}
We note that an upper limit to the achievable $\beta$ values in graphene is set by the condition that
potential energy change over a magnetic length should be less then the energy difference between adjecent
LLs to avoid tunneling along the parallel electric field.
For the $n=0,1$ Landau levels, this means
$e E_\parallel\ell<\epsilon_1^\text{G} - \epsilon_0^\text{G} = \frac{\hbar v\sqrt2}{\ell}(1-\beta^2)^{3/4}$,
which sets $\beta<0.75$.

In \BEDT\ and potentially other quasi-2D organic materials with massless Dirac cones,\cite{theta,theta2,Choji2011}
on the other hand, a quantitative analysis of this fingerprint,
in combination with other information on Fermi velocities if the tilt turns out to be large,
helps us determine the tilting parameters of the massless Dirac cones.
We notice finally that the same magneto-optical features should appear in strained graphene,
where uniaxial strain yields tilted Dirac cones \cite{Goerbig} that are predicted to show
a clear signature also in Raman spectroscopy.\cite{sonia}

\begin{acknowledgments}
We acknowledge support from the Hungarian Scientific Research Funds No.\ K105149.
C.\ T.\ was supported by the Hungarian Academy of Sciences.
Supercomputer facilities were provided by National Information Infrastructure Development Institute,
Hungary. J. S. was partially supported
by Campus France supervised by the French Ministry of Education, and
also acknowledges hospitality from Laboratoire de Physique des Solides of Universit\'e Paris-Sud.
We thank M.\ Monteverde, C.\ Faugeras, M.\ Orlita, M.\ Potemski, A. P\'alyi and
K.\ Kamar\'as for useful discussions.
\end{acknowledgments}

\appendix

\section{Derivation of the minimal Weyl Hamiltonian}
\label{derivation}

\begin{figure}[htbp]
\begin{center}
\includegraphics[height=\columnwidth,angle=270,keepaspectratio]{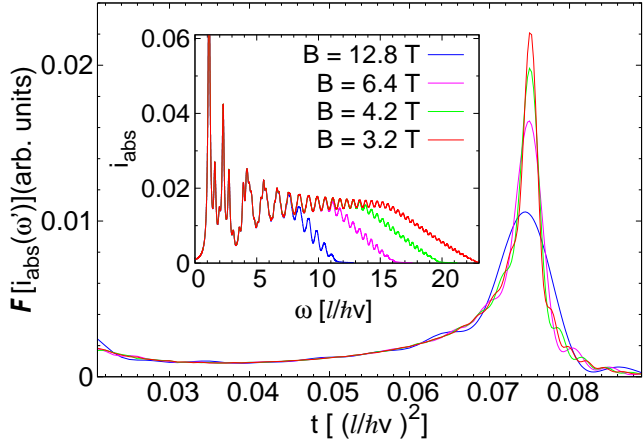}
\end{center}
\caption{\label{cutoffdependence}
(Color online)
The one-sided Fourier transform of $i_\text{abs}^\text{W}(\omega)$ for the Weyl Hamiltonian for linearly polarized light
in the tilt direction [Eq.~(\ref{tilteq})] and $\beta = 0.5$  for various Landau level cutoff values:
25, 50, 100, 150, which corresponds to $B = 12.8, 6.4, 4.2, 3.2 $ T, respectively.
Inset: the original $i_\text{abs}^\text{W}(\omega)$ function.
}
\end{figure}

Massless Dirac carriers of \BEDT\ are described by the minimal Weyl Hamiltonian \cite{kobayashi} 
using four parameters $v_0^x, v_0^y, v_x, v_y$ as follows
\beq\label{eq:HamKobayashi}
\hat{H}_\text{W}= v_x \hat p_x \sigma_x + v_y\hat p_y \sigma_y + (v^x_0\hat p_x + v^y_0\hat p_y )\bone.
\eeq
The inclination of the Dirac cone is determined by the combined effect of the tilt and the anisotropy. 
We characterize the anisotropy of the Dirac cones by the quotient $v_x / v_y$. 
Even if $v_x=v_y$, the constant energy slices are not concentric because $(v_0^x,v_0^y)\neq(0,0)$.
We quantify this tilt by the parameter
\begin{equation}
\eta = \sqrt{(v_0^x / v_x)^2 + (v_0^y /v_y)^2}. 
\end{equation}

Following Refs.~\onlinecite{Morinari} and \onlinecite{Sari}, we use a rescaled and rotated coordinate system,
defined by the transformation
\begin{gather}
\label{mapy}
\begin{pmatrix}
\tilde x \\ \tilde y
\end{pmatrix}
=\begin{pmatrix}
\cos\phi & \sin\phi \\
-\sin\phi & \cos\phi
\end{pmatrix}
\begin{pmatrix}
x \\ y \frac{v_x}{v_y}
\end{pmatrix},\\
\label{mapky}
\begin{pmatrix}
\tilde p_x \\ \tilde p_y
\end{pmatrix}
=\begin{pmatrix}
\cos\phi & \sin\phi \\
-\sin\phi & \cos\phi
\end{pmatrix}
\begin{pmatrix}
\hat p_x \\ \hat p_y\frac{v_y}{v_x}
\end{pmatrix},
\end{gather}
in terms of the rotation angle $\phi$.
Rescaling $\hat p_y$ removes the anisotropy, and the rotation brings $\tilde p_x$ in the tilt direction if we choose
\begin{equation}
\label{tilteq}
\cos\phi = \frac{v_0^x v_y}{\sqrt{(v_0^y v_x)^2 + (v_0^x v_y)^2}}.
\end{equation}
Then the four-parameter Hamiltonian $\hat{H}_\text{W}$ can be rewritten in a simpler form
\begin{equation}
\label{Htilted}
\hat H_\text{W}=  v_x \tilde{\bp} \cdot \sigmab + v_0 \tilde{p}_x \bone,
\end{equation}
where $\tilde{\bp}$ is measured from the direction of the tilt, and
\begin{equation}
\label{vnull}
v_0 =\eta v_x.
\end{equation}
The Peierls substitution in Eq.~(\ref{Htilted}) then yields Eq.~(\ref{eq:HamOrg}).

\section{Determination of the cutoff}
\label{detcutoff}

As the number of Landau levels within the range of validity of the Weyl Hamiltonian is limited,
we examine the robustness of our results against the change of the Landau level cutoff.
Previously, we estimated the number of available Landau levels is $300/B_\perp$.\cite{Sari}
Fig.~\ref{cutoffdependence} shows the effect of an arbitrary change of the cutoff.
The oscillatory behavior remains visible in the $i_\text{abs}^\text{W}(\omega)$ curves even at low cutoffs.
The relevant peaks of $\mathcal{F}[i_\text{abs}(\omega')]$ also survive, though with reduced height.
We can see that the oscillatory behavior of $i_\text{abs}^\text{W}$ is present for as few as 50 Landau levels,
which is a safe choice for reasonable magnetic fields.


\begin{thebibliography}{99}
\bibitem{foot} The study of this effect is delegated to future work.
\end{thebibliography}
\end{document}